\documentclass[conference]{IEEEtran}
\usepackage{cite}
\usepackage{amsmath,amssymb,amsfonts}
\usepackage{graphicx}
\usepackage{tabularx}
\usepackage{color}
\usepackage{fancyvrb}
\usepackage{fvextra}
\usepackage{float}
\usepackage[table]{xcolor}
\usepackage{tcolorbox}
\usepackage{minted}
\usepackage{cuted}
\usepackage{algorithmicx}
\usepackage[noend]{algpseudocode} 
\usepackage{mdframed}
\usepackage{multicol}
\usepackage{multirow}
\usepackage{textcomp}
\usepackage{booktabs}
\usepackage{url}

\usepackage[T1]{fontenc}
\usepackage{newtxtext,newtxmath}

\begin{document}
\title{LLM-assisted gNB Parameter Configuration for Radio Access Networks}
\author{\IEEEauthorblockN{
        Yao-Cong Dong,\IEEEauthorrefmark{1}\IEEEauthorrefmark{4}
        Maria Amparo Canaveras Galdon, \IEEEauthorrefmark{2}\IEEEauthorrefmark{3}\IEEEauthorrefmark{4}
        Ari Uskudar,\IEEEauthorrefmark{2}
        Kuntal Chowdhury, \IEEEauthorrefmark{2}\\
        Edwin K. P. Chong,\IEEEauthorrefmark{3}
        and
        Ray-Guang Cheng\IEEEauthorrefmark{1}\\
  }
    \IEEEauthorblockA{\IEEEauthorrefmark{1}
    Dept. of Electronic and Computer Engineering, National Taiwan University of Science and Technology, Taiwan
    }  \IEEEauthorblockA{\IEEEauthorrefmark{2}
    NVIDIA, USA 
    }
    \IEEEauthorblockA{\IEEEauthorrefmark{3}
Colorado State University, USA  \\
    }
    \IEEEauthorblockA{\IEEEauthorrefmark{4}
These authors contributed equally to this work.  \\
    }
    Email: acanaveras@nvidia.com, crg@mail.ntust.edu.tw
    }

\maketitle


\begin{abstract}
gNB parameter misconfigurations are a common cause of system failures in radio access networks (RANs), and their diagnosis and correction rely on manual analysis of complex network logs that does not scale well. This paper proposes a large language model (LLM)-assisted framework for automatic gNB parameter configuration. The framework adopts a synthetic data generation pipeline following a configuration–log–correction workflow. Starting from a workable configuration and the gNB technical references, the pipeline uses a commercial LLM to generate modified configurations and derive structured reasoning traces from gNB error logs. The synthetic training data maps network states to corrective actions and is used to fine-tune an LLM for configuration correction. During inference, the fine-tuned LLM generates valid and deployable gNB parameter configurations from gNB error logs. The framework is validated on an OpenAirInterface (OAI) gNB testbed with 480 unseen misconfiguration scenarios, where fine-tuning improves correction accuracy from 13.8\% (zero-shot baseline) to 85.4\%, and retrieval-augmented generation (RAG) further improves accuracy to 92.7\%. The results demonstrate that the framework may enable automated recovery from misconfigurations without manual intervention and supports scalable and autonomous RAN operation.

\end{abstract}


\section{Introduction}

gNB parameter misconfigurations are a major source of system failures in radio access networks (RANs). In disaggregated 5G gNB, the Central Unit (CU) and Distributed Unit (DU) must be configured properly to ensure correct network operation. These network elements generate event logs that record signaling procedures, parameter changes, and error reports during system operation. Misconfigurations or dynamic network conditions often lead to failures that are reflected in these logs. However, these logs rarely point directly to the root cause. 
For example, a single incorrect parameter, such as a mismatched cell identity, an out-of-range transmission power, or an invalid frequency offset, can prevent F1 interface establishment or UE attachment. The resulting logs provide only indirect error indications, making diagnosis and correction reliant on manual analysis. Manual diagnosis and reconfiguration are time-consuming, require expert knowledge, and do not scale well with network size and complexity. Automated configuration correction is therefore critical for enabling AI-driven and autonomous RAN operation. 

Existing approaches for network automation remain limited in addressing this problem. Traditional Self-Organizing Networks (SON) and rule-based fault management rely on predefined logic and static rules. These methods are effective for previously characterized fault conditions. However, they struggle to adapt to increasing network complexity, unseen failure modes, and interdependent cross-module issues.
Machine-learning-based methods for anomaly detection and key performance index (KPI) optimization are based on labeled datasets and structured feature representations. Such reliance limits their applicability to raw, noisy, and unstructured system logs in real-world environments.
In practice, gNB logs contain heterogeneous and context-dependent information, such as signaling traces, timing errors, and module interactions. Effective interpretation of gNB logs requires semantic understanding and causal reasoning. This challenge is compounded in multi-vendor O-RAN deployments, where log formats and semantics vary across vendors, further complicating automated interpretation. Current network automation frameworks lack this reasoning capability and cannot derive actionable configuration adjustments directly from log text. As 5G networks evolve toward dynamic, multi-vendor, and software-defined environments, this limitation increasingly constrains autonomous network management. 
Large language models (LLMs) offer a promising direction. They can comprehend unstructured text, perform multi-step reasoning, and generate actionable recommendations. Recent advances in LLM reasoning have shown strong performance in domains such as mathematics and coding, where high-quality training data with structured reasoning traces and verifiable outputs are available~\cite{gunasekar2023textbooks}.
However, network troubleshooting has not yet benefited from comparable reasoning advances. This gap arises because the training data that drive success in mathematics and coding do not exist in this domain. Raw gNB logs are rarely paired with expert diagnostic rationales, and proposed configuration fixes cannot automatically be verified for correctness. 

To bridge this gap, this paper proposes an LLM-assisted framework for automatic gNB parameter configuration. The framework adapts structured reasoning techniques originally developed for mathematics and coding to the interpretation of gNB event logs. It uses synthetic data grounded in real testbed execution to generate the training signal that the domain naturally lacks. The main contributions of this paper are as follows:
\begin{itemize}
    \item A reproducible configuration mutation pipeline that uses an LLM to systematically generate realistic configuration mutations and deploys every mutation against a real OAI 5G system, capturing error logs with known ground truth to ensure all training data reflects genuine failure behavior.
    \item A synthetic data generation pipeline that maps gNB error logs to structured reasoning traces via three distinct prompt strategies (deductive chain, systematic elimination, iterative self-refinement).
    \item An SFT training pipeline leveraging diverse reasoning traces to instill domain-specific, vendor-specific, and software-version-specific network diagnostic knowledge. network diagnostic knowledge in an LLM, evaluated under both RAG and non-RAG data conditions.
    \item The synthetic dataset, training pipeline, and evaluation code are open-sourced to enable reproducibility and community adoption.\footnote{\url{https://github.com/bmw-ece-ntust/gNB-error-configuration}}
\end{itemize}

The rest of this paper is organized as follows.
Section~\ref{sec:related} reviews related work.
Section~\ref{sec:system_model} defines the system model.
Section~\ref{sec:framework} presents the LLM-assisted framework, comprising configuration mutation generation, reasoning trace synthesis, and supervised fine-tuning.
Section~\ref{sec:results} presents the experimental results.
Section~\ref{sec:conclusion} provides concluding remarks.

\section{Related Work}
\label{sec:related}


The O-RAN ALLIANCE architecture provides a natural foundation for AI-driven network management.
Polese {\it et al.}~\cite{polese2023understanding} present a comprehensive tutorial on O-RAN, detailing the disaggregated architecture, open interfaces (O1, E2, A1), and the Service Management and Orchestration (SMO) framework that enables data-driven closed-loop control through RAN Intelligent Controllers (RICs). Our system model (Fig.~\ref{fig:model}) adopts the O-RAN disaggregated architecture with an SMO-mediated management platform, but focus specifically on the diagnostic reasoning task—using a fine-tuned LLM as the operator that analyzes error logs collected via standardized interfaces to infer configuration corrections.

Recent work has explored LLMs for telecommunications and network management~\cite{llm4telecom_survey}.
While general-purpose LLMs excel in natural language processing, they often struggle with the intricate details of telecommunications standards. For instance, even state-of-the-art models like GPT-4 face failure rates approaching 50\% when handling queries related to 3GPP technical specifications \cite{zou2025telecomgpt}, further underscoring the critical necessity for specialized domain-adaptation pipelines. To address these limitations, several research efforts have focused on enhancing LLMs' technical proficiency within the sector.
Zhou {\it et al.}~\cite{llm4telecom_survey} survey LLMs for telecommunications covering network planning, optimization, and troubleshooting.
Maatouk {\it et al.}~\cite{llm_network_modeling} discuss domain adaptation and highlight the need for telecom-specific training data.
Lin {\it et al.}~\cite{lin2025fd} propose FD-LLM for fault diagnosis, while Rezaei {\it et al.}~\cite{rezaei2025fedllmguard} develop a federated LLM for 5G anomaly detection.
However, none describe a reproducible pipeline for creating network diagnostic training data grounded in real system behavior.


Ouyang {\it et al.}~\cite{ouyang2022training} established supervised fine-tuning (SFT) as a foundational technique for aligning LLMs with desired behavior by training on curated demonstration data, enabling smaller models to replicate capabilities of larger teachers.
Chain-of-Thought prompting~\cite{wei2022chain} showed that step-by-step reasoning dramatically improves LLM performance.
STaR~\cite{zelikman2022star} generates rationales with the answer provided, then fine-tunes on (problem~$\to$~rationale~$\to$~answer) pairs.
Lightman {\it et al.}~\cite{lightman2024lets} introduced process reward models scoring intermediate steps.
Self-Refine~\cite{madaan2023selfrefine} uses iterative generate--critique--refine cycles.
For training data, Gunasekar {\it et al.}~\cite{gunasekar2023textbooks} showed synthetic data can match real data for specialist models, and Duanis {\it et al.}~\cite{duanis2025json} found generating only necessary JSON modifications is more stable than complete regeneration.
Our work builds directly on these foundations: we use SFT~\cite{ouyang2022training} to distill diagnostic reasoning from a large teacher model into a compact 32B-parameter model; CoT prompting~\cite{wei2022chain} structures every reasoning trace as a step-by-step analysis; and we design three complementary prompt templates for synthetic trace generation, each grounded in a different reasoning paradigm (Section~\ref{sec:trace_gen}).
Following Gunasekar {\it et al.}~\cite{gunasekar2023textbooks}, all training data is synthetically generated, and we adopt the modification-only generation principle of Duanis {\it et al.}~\cite{duanis2025json} for our configuration mutation pipeline.

\section{System Model}
\label{sec:system_model}

\begin{figure}[h]
\centering
\includegraphics[width=0.85\linewidth]{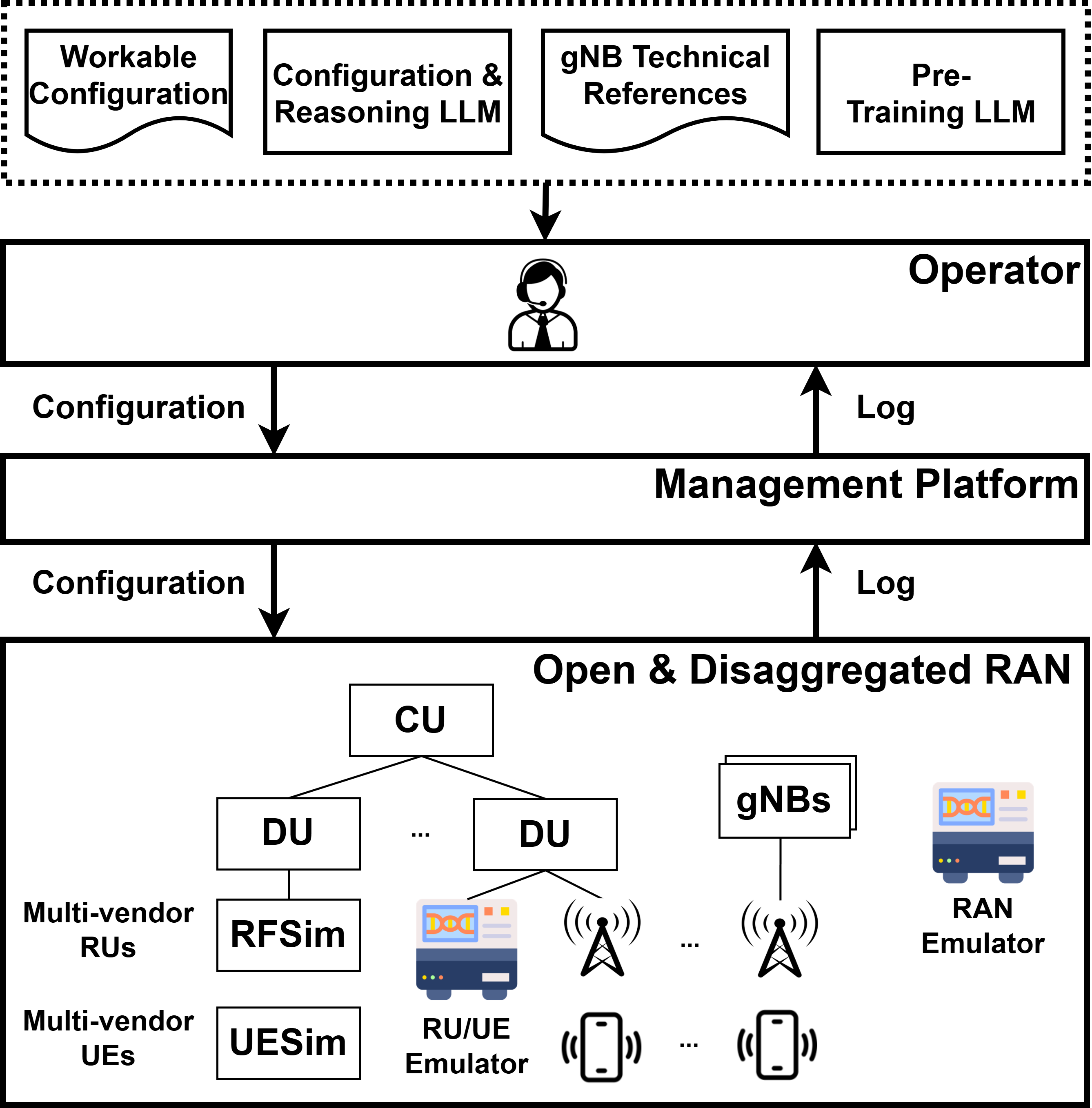}
\caption{System model of the O-RAN management and control architecture.}
\label{fig:model}
\end{figure}

Figure~\ref{fig:model} shows the system model of the O-RAN management and control architecture considered in this paper.
The system consists of three main components: an operator, a management platform, and an open and disaggregated RAN.
The operator may be a human, an automated system, or an LLM equipped with RAN-specific domain knowledge, including RAN terminology and log semantics. 

The operator is assumed to acquire background knowledge of gNB parameter configuration from a workable(golden) configuration and the gNB technical documentation references such as vendor manuals or source code. The operator can leverage an LLM for configuration and reasoning, or further enhance a pre-training LLM to support decision making. With this knowledge, the operator sends configuration parameters to the gNB and collects logs through the management platform. In practice, operators also collect operational measurements (OMs), key performance indicators (KPIs), and platform-specific telemetry for performance analysis. The current work focuses on event logs as the primary diagnostic input; incorporating OMs and KPIs is left for future work. The management platform may be an Operations Support System (OSS) or a Service Management and Orchestration (SMO) defined by the O-RAN ALLIANCE. Through this platform, the operator configures RAN components and collects operational logs from the open and disaggregated RAN using the O1 interface and YANG models specified by 3GPP~\cite{TS128541_NRM}.
In this paper, we propose an LLM-assisted framework to implement the operator. The framework replaces manual diagnosis with an automated reasoning pipeline and generates corrected gNB parameter configurations from error logs.

The performance of the operator is evaluated using two criteria: parameter identification accuracy and correction quality. Parameter identification accuracy measures whether the operator correctly identifies the misconfigured parameter(s). For multi-parameter scenarios, the accuracy is computed on a per-parameter basis. Correction quality evaluates whether the operator produces a valid correction with the correct parameter name and value. A graded scale is used to distinguish exact matches, format-variant matches, partially correct identifications, and incorrect outputs. For an LLM-based Operator, these criteria are assessed using Exact-Match accuracy and LLM-as-Judge scoring, as detailed in Section~\ref{sec:framework}.

\section{LLM-assisted Framework for gNB Parameter Configuration}
\label{sec:framework}

\begin{figure}[t]
    \centering
    \includegraphics[width=\linewidth]{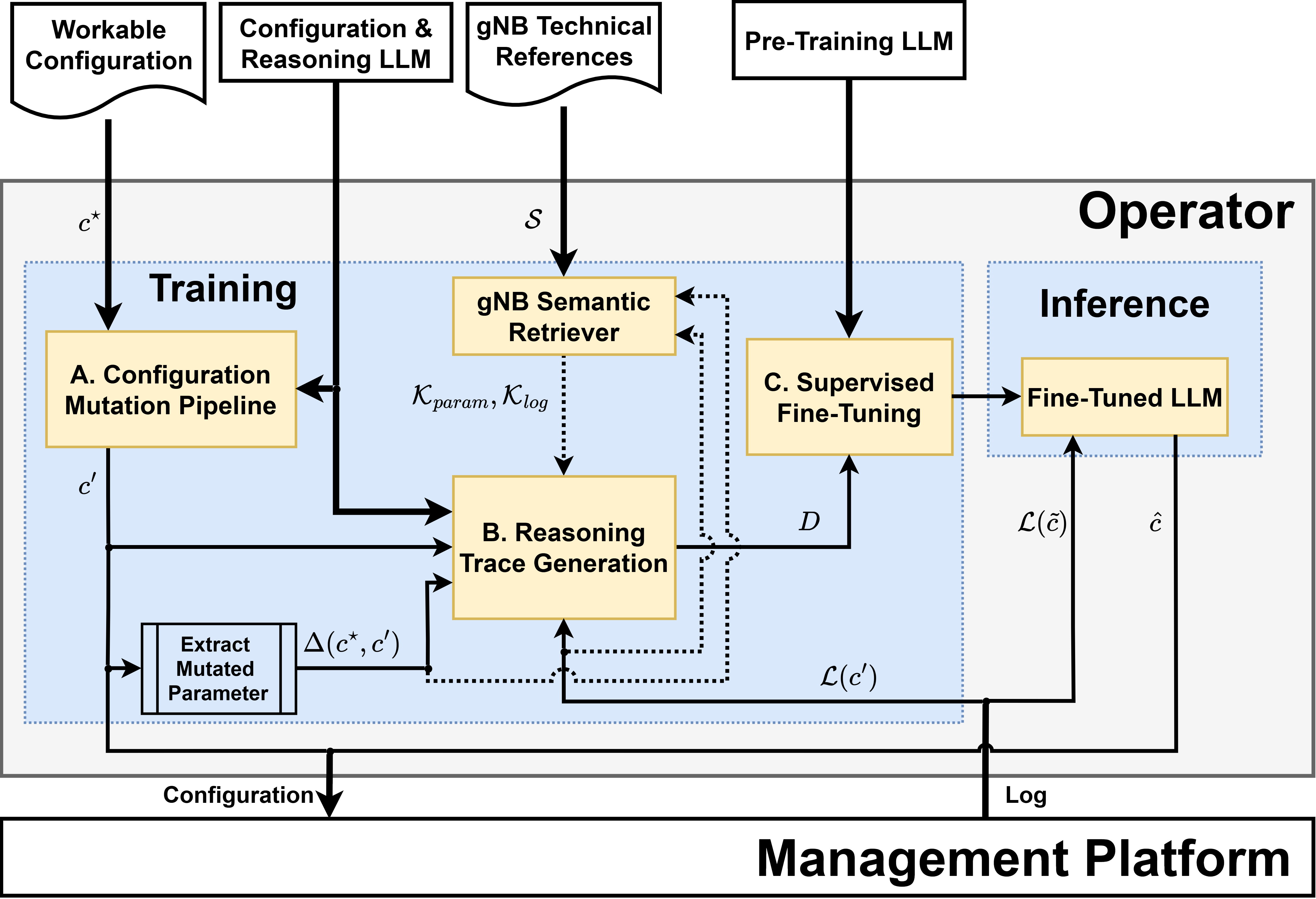}
    \caption{Functional block diagram of the proposed LLM-assisted framework
    for gNB parameter configuration.}
    \label{fig:framework}
\end{figure}

Figure~\ref{fig:framework} shows the functional block diagram of the proposed
LLM-assisted framework for gNB parameter configuration. The framework consists
of two main blocks: a training block and an inference block. The training block
executes three sequential stages: (A)~a configuration mutation pipeline that
generates erroneous configurations and collects the corresponding error logs
from the RAN; (B)~a reasoning trace generation stage that produces structured
diagnostic reasoning traces from those logs; and (C)~a supervised fine-tuning
stage that uses the resulting dataset to fine-tune a pre-training LLM. The
fine-tuned model constitutes the inference block, which takes gNB error logs
as input and generates corrected parameter configurations for deployment
through the management platform.
Let $\mathcal{C}$ denote the configuration space of the disaggregated RAN;
$c^\star \in \mathcal{C}$ be a given workable configuration under which the
RAN operates correctly; $c' \in \mathcal{C}$ be a modified configuration
generated by applying mutations to $c^\star$, though such mutations do not
necessarily lead to system errors; $\tilde{c} \in \mathcal{C}$ be an erroneous
configuration that leads to abnormal behavior of the RAN and generates error
logs; and $\Delta(c^\star, c')$ denote the set of configuration variables that
differ between $c^\star$ and $c'$. Let $\mathcal{L}(\tilde{c})$ denote the logs generated when the RAN operates under the erroneous configuration $\tilde{c}$. Given $\tilde{c}$ and the observed logs $\mathcal{L}(\tilde{c})$, the operator aims to infer a corrected configuration $\hat{c} \in \mathcal{C}$ that restores proper RAN operation. Let $L_{pre}$ denote the pre-training LLM; $L_{conf}$ the Configuration LLM responsible for generating mutations from $c^\star$; and $L_{reason}$ the Reasoning LLM responsible for generating diagnostic traces and, after fine-tuning, predicting~$\hat{c}$.

Let $\mathcal{S}(x)$ denote the reference to technical documentation of an implementation of a gNB implementation $x$—where $x \in \mathcal{X}$ and $\mathcal{X}$ represent the set of available software platforms (e.g. O-RAN Software Community, OpenAirInterface, srsRAN, or commercial solutions); since passing $\mathcal{S}{(x)}$ in its entirety as input to LLM is impractical, the framework incorporates an RAG mechanism. In this mechanism, $\mathcal{K}_{log}$ and $\mathcal{K}_{param}$ represent the log and parameter code chunks extracted from $\mathcal{S}{(x)}$, corresponding to $\mathcal{L}(\tilde{c})$ and $\Delta(c^\star, \tilde{c})$, respectively. Let $D$ denote the training dataset composed of $T$ paired with their respective input contexts. The following subsections describe the implementation of each training stage.
\subsection{Configuration Mutation Pipeline}
The configuration mutation pipeline generates a large number of erroneous
configurations $\tilde{c}$ from a workable configuration $c^\star$ using the $L_{conf}$. The strategy follows two principles: (i) decomposing generation tasks into local subproblems to improve accuracy~\cite{wei2022chain}, and (ii) applying minimal parameter changes, which is more stable than full regeneration~\cite{duanis2025json}. A mutation prompt $P_{\text{mut}}$ guides $L_{conf}$, which receives $c^\star$---provided in both its native \texttt{.conf} format and a structured JSON representation---as input, analyzes its structure and modifiable parameters,
and produces for each test case a delta record specifying the modified key,
its original value, and the injected error value. A patching script then
applies each delta to $c^\star$ at the field level to produce $\tilde{c}$.
After generation, duplicate mutations targeting the same parameter and value
are removed and the number of cases per mutation target is normalized to
ensure balanced experimental conditions across fault categories.
Each $\tilde{c}$ is then deployed to the RAN via the management platform, which dynamically pairs the mutation target (CU or DU) with $c^\star$ following a 20-second sequential start for the CU, DU, and UE. The platform collects the resulting logs, performs environment cleanup, and filters for UE attachment failures to construct the dataset $\{\tilde{c},\, \mathcal{L}(\tilde{c}),\, \Delta(c^\star, \tilde{c})\}$. Current focus on UE attachment failures provides a clear diagnostic baseline, leaving KPI-based performance analysis for future research. Only configurations confirmed to produce observable failures are retained, ensuring that $c' = \tilde{c}$ throughout the dataset. These outputs form the inputs to the subsequent reasoning trace generation stage.

\subsection{Reasoning Trace Generation}
\label{sec:trace_gen}
In this stage, we integrate multiple prompt engineering strategies into a
reasoning trace prompt---defining the role, reasoning steps, and output
structure---to guide $L_{reason}$ in generating diagnostic processes with
engineering logic.

To evaluate the impact of implementation-level knowledge, we generate reasoning traces under two distinct conditions: $T_{\text{base}}$ and $T_{\text{rag}}$. In the baseline scenario, $T_{\text{base}}$ is generated where $L_{\text{reason}}$ relies solely on its pre-training knowledge to infer how $\Delta(c^\star, \tilde{c})$ cause the observed $\mathcal{L}(\tilde{c})$ from the input tuple $\{\tilde{c}, \mathcal{L}(\tilde{c}), \Delta(c^\star, \tilde{c})\}$. Under the RAG condition, $T_{\text{rag}}$ is produced by augmenting prompts with code chunks retrieved from $\mathcal{S}{(x)}$, expanding the input set to $\{\tilde{c}, \mathcal{L}(\tilde{c}), \Delta(c^\star, \tilde{c}), \mathcal{K}_{\text{log}}, \mathcal{K}_{\text{param}}\}$ and enabling the model to ground its causal analysis in actual implementation details.While retrieval errors cause flawed reasoning, accurate labels enable SFT’s implicit noise tolerance, ensuring diagnostic robustness.

To maximize reasoning diversity, we employ three distinct prompt strategies: (i) \textbf{Deductive Chain} (STaR-based~\cite{zelikman2022star}) follows the logic of Observations $\to$ Analysis $\to$ Correlation $\to$ Hypothesis $\to$ Fix ; (ii) \textbf{Systematic Elimination} (PRM-inspired~\cite{lightman2024lets}) proceeds through Symptom Collection $\to$ Differential Diagnosis $\to$ Systematic Elimination $\to$ Final Diagnosis $\to$ Fix ; and (iii) \textbf{Iterative Self-Refinement} (Self-Refine~\cite{madaan2023selfrefine}) involves an Evidence Scan $\to$ Initial Hypothesis $\to$ Verification $\to$ Refinement $\to$ Convergence $\to$ Fix cycle. All three strategies are applied independently to each training case to diversify diagnostic patterns during SFT ; however, inference yields a single output without requiring strategy selection.

\subsection{Supervised Fine-Tuning}
\label{sec:sft}

\subsubsection{Data Preparation}
To construct the $D$, we aggregate the $T$ generated under the two aforementioned knowledge-grounding conditions. Specifically, each entry in $D$ is formalized according to its source: under the RAG-enhanced condition, the entry is a tuple $\{\tilde{c}, \mathcal{L}(\tilde{c}), \Delta(c^\star, \tilde{c}), \mathcal{K}_{\text{log}}, \mathcal{K}_{\text{param}}, T_{\text{rag}}\}$, where $T_{\text{rag}}$ reflects logic grounded in source code; whereas under the baseline condition, it is formalized as $\{\tilde{c}, \mathcal{L}(\tilde{c}), \Delta(c^\star, \tilde{c}), T_{\text{base}}\}$.  These entries form the dataset used for fine-tuning the model to diagnose and resolve various gNB failure scenarios.
The synthetic dataset comprises 2,400 unique misconfiguration cases covering both CU and DU components. All 2,400 deployed mutations yielded observable UE attachment failures, confirming $c'$ =  $\tilde{c}$ across the complete dataset.
An 80/20 problem-level split yields 1,920 training and 480 test scenarios with zero overlap between splits.
Each training case is presented under all three prompt styles, yielding a theoretical maximum of $1{,}920 \times 3 = 5{,}760$ traces.
Both the No-RAG and RAG SDG pipelines apply a quality filter that removes individual traces where $L_{\text{reason}}$ produced an incomplete JSON fix for multi-parameter cases---i.e., the fix contained fewer keys than the number of misconfigured parameters.
Single-parameter cases always pass this filter.
After filtering, 118 multi-parameter cases retain only 2 traces (the systematic elimination prompt consistently produced incomplete fixes for these cases), while the remaining 1,802 cases retain all~3.
This yields $1{,}802 \times 3 + 118 \times 2 = 5{,}642$ training traces per data condition.
The same filtering is applied identically to both the No-RAG and RAG pipelines, ensuring a fair comparison.
Table~\ref{tab:dataset} summarizes the dataset statistics.

\begin{table}[t]
    \centering
    \caption{Dataset Statistics}
    \label{tab:dataset}
    \begin{tabular}{ll}
        \toprule
        \textbf{Property} & \textbf{Value} \\
        \midrule
        Total scenarios (80/20 split) & 2,400 (1,920 train / 480 test) \\
        Train: single / multi-param & 988 / 932 \\
        Test: single / multi-param & 252 / 228 \\
        Prompt styles per condition & 3 \\
        SFT traces per condition & 5,642 ($1{,}920{\times}3 - 118$ filtered) \\
        \bottomrule
    \end{tabular}
\end{table}

\subsubsection{Training}
We select QwQ-32B~\cite{qwq32b} as the {$L_{pre}$}, a 32B-parameter reasoning model built on the Qwen2.5 architecture with reinforcement-learning-based post-training for enhanced reasoning.
SFT on $D$ produces the fine-tuned model.
Training uses the NeMo-RL framework with Megatron backend~\cite{shoeybi2019megatron} and runs for 3 epochs.
Table~\ref{tab:hyperparams} lists the training hyperparameters.
\begin{table}[t]
    \centering
    \caption{SFT Training Hyperparameters}
    \label{tab:hyperparams}
    \begin{tabular}{ll}
        \toprule
        \textbf{Parameter} & \textbf{Value} \\
        \midrule
        Base Model & QwQ-32B \\
        Framework & NeMo-RL + Megatron \\
        GPUs & 8$\times$ H100 (80\,GB) \\
        Learning Rate & $1 \times 10^{-5}$ (constant) \\
        Global Batch Size & 32 \\
        Max Sequence Length & 4,096 tokens \\
        Precision & bfloat16 \\
        Optimizer & AdamW ($\beta_1{=}0.9$, $\beta_2{=}0.98$) \\
        Epochs & 3 \\
        Sequence Packing & Enabled \\
        \bottomrule
    \end{tabular}
\end{table}

\subsubsection{Model Evaluation}
The fine-tuned model is evaluated on the held-out test set using the performance 
metrics defined in Section~\ref{sec:system_model}, implemented as follows. 
\textbf{Exact-Match (EM) accuracy} compares the parameter name extracted from 
the model's JSON fix output against the ground-truth field —defined as the original parameter name(s) and value(s) in c*, known from $\Delta(c^\star, \tilde{c})$— using key-name matching 
at any JSON depth. For multi-parameter problems, EM is micro-averaged (each 
parameter scored independently). \textbf{LLM-as-Judge scoring} uses 
Qwen2.5-32B-Instruct at temperature~0.0 for deterministic scoring with a 0--3 
rubric: 3~=~exact match (name and value correct); 2~=~correct parameter, different 
format; 1~=~related but wrong parameter; 0~=~not inferable or no JSON fix produced. 
Accuracy is defined as the fraction of samples scoring $\geq 2$.

\section{Experimental Results}
\label{sec:results}

\subsection{Experimental Setup}

This experimental environment utilizes Open5GS as the core network and OAI gNB (CU/DU) and OAI UE, employing RF simulator mode to implement an Open \& Disaggregated RAN architecture. All RAN components are deployed as containerized units within a kubernetes environment. Furthermore, the $\mathcal{S}(\text{OAI})$ is provided as input to the gNB Semantic Retriever to facilitate the comparative experiments between RAG and No-RAG. The gNB parameters are configured for the n78 band (3.5 GHz TDD), 106 PRBs, and subcarrier spacing of 30 kHz (numerology 1). Since the primary objective is to collect errors originating from the gNB itself, the configuration between the Open5GS and the OAI UE was verified to be correct prior to the experiments. Table~\ref{table:models} lists the frameworks and models used throughout the pipeline.
All experiments run on a single node with 8$\times$ NVIDIA H100 (80\,GB) GPUs.
SDG inference uses vLLM~\cite{kwon2023vllm} with tensor parallelism (TP=8).

\begin{table}[t]
    \centering
    \caption{Frameworks and Models Used}
    \label{table:models}
    \begin{tabular}{ll}
        \toprule
        \textbf{Module} & \textbf{Framework / Modle} \\
        \midrule
        Mutation Generator & NeMo Skills / GPT-OSS-120B \\
        Reasoning Trace Generator & NeMo Skills / GPT-OSS-120B \\
        Semantic Retriever (RAG) & NeMo Retriever / Nemotron Embedding \\
        SFT Training & NeMo RL / QwQ-32B~\cite{qwq32b} \\
        Evaluation Judge & vLLM / Qwen2.5-32B-Instruct~\cite{zheng2024llmjudge} \\
        \bottomrule
    \end{tabular}
\end{table}
\subsection{Evaluation Results}

\begin{figure}[t]
    \centering
    \includegraphics[width=0.85\linewidth]{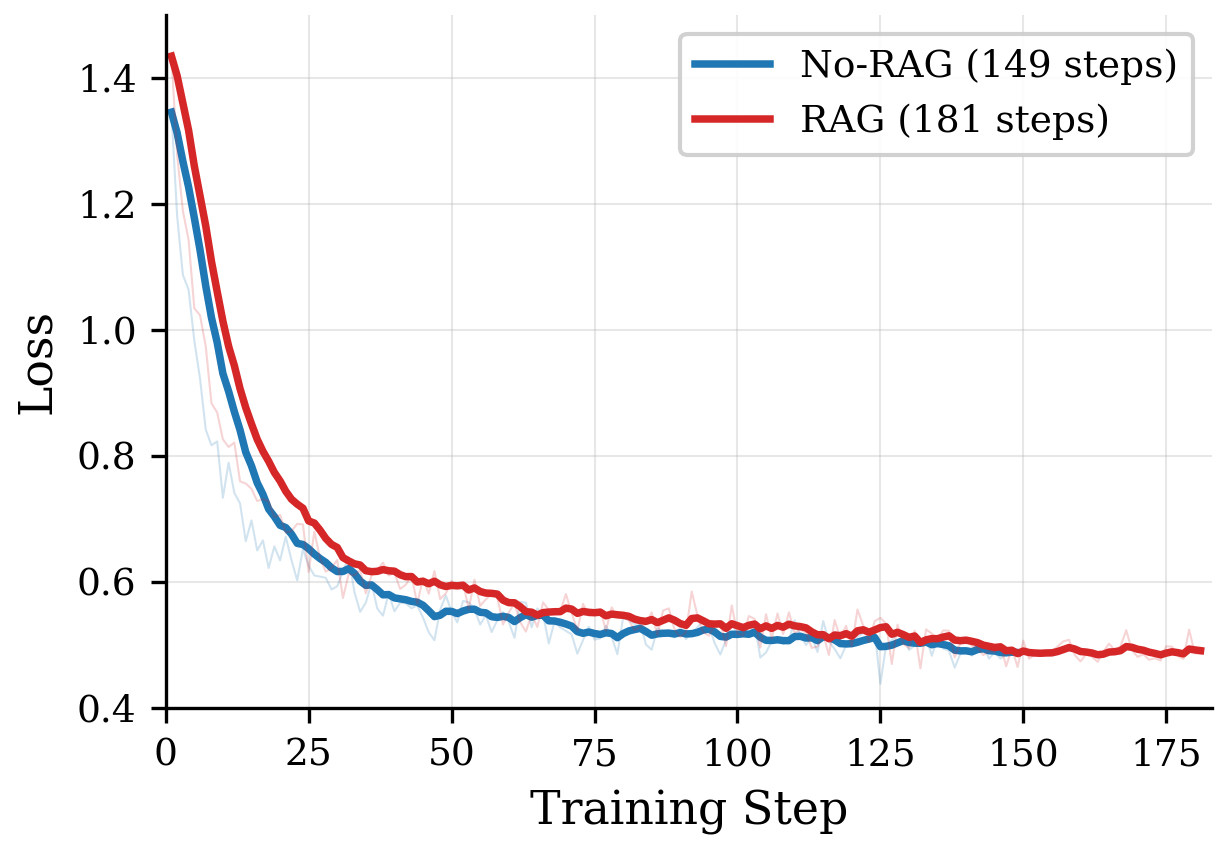}
    \caption{SFT training loss for QwQ-32B under both data conditions (lr=$1{\times}10^{-5}$, batch=32). Solid lines show exponential moving average ($w{=}9$); shaded regions show raw loss.}
    \label{fig:training_loss}
\end{figure}

We evaluate the fine-tuned models on 480 held-out test scenarios under three conditions: zero-shot baseline, SFT (No-RAG), and SFT (RAG).
Figure~\ref{fig:training_loss} shows that both SFT tracks converge to comparable final loss ($\approx$0.49), confirming that the additional source-code context in the RAG condition does not impede optimization.
SFT dramatically improves performance over the zero-shot baseline across all metrics.
The No-RAG SFT model achieves 85.4\% exact-match accuracy and 89.4\% accuracy (score~$\geq 2$), representing a 71.6 percentage-point improvement in EM over the baseline.
The RAG SFT model achieves 75.4\% EM but a higher accuracy of 92.7\%, indicating that RAG-augmented traces lead the model to identify correct parameters more often but sometimes in different formats (e.g., using an absolute JSON path \texttt{gNBs.[0].nr\_cellid} instead of the relative key \texttt{nr\_cellid}, or hexadecimal values like \texttt{0x64} instead of \texttt{100}).
The baseline model fails to produce a usable JSON fix for 73.1\% of problems (score~0), reflecting the difficulty of the task without domain-specific training.
Notably, the RAG model produces significantly more score-2 results (83 vs.\ 19), explaining the accuracy gap despite lower EM---it identifies the correct parameter but often expresses the fix using alternative JSON path formats or value representations learned from the source code context.

\begin{figure}[t]
    \centering
    \includegraphics[width=0.85\linewidth]{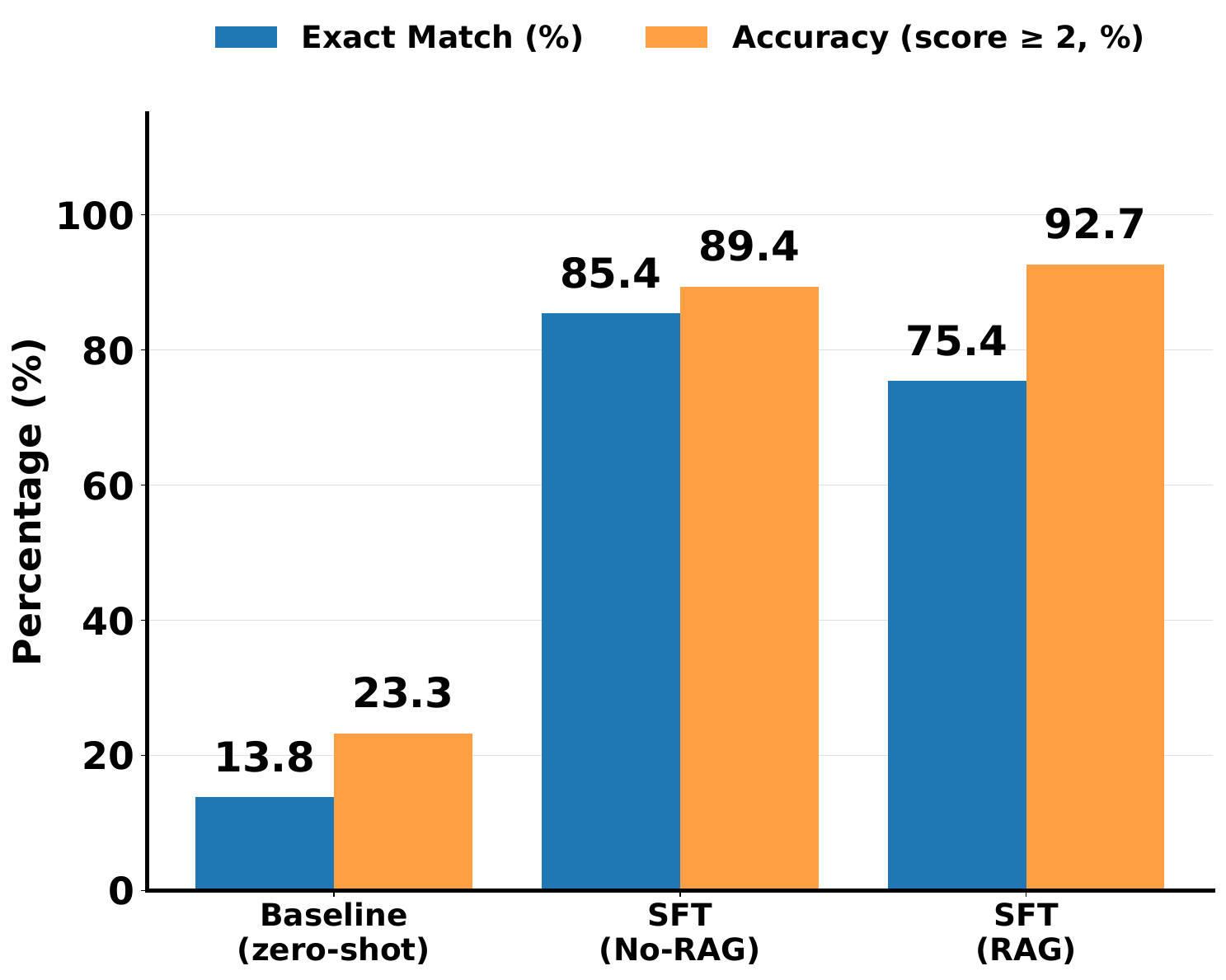}
    \caption{Comparison of exact-match accuracy and accuracy (score $\geq 2$) across experimental conditions.}
    \label{fig:results}
\end{figure}

The results demonstrate that the proposed synthetic data pipeline produces training data of sufficient quality to dramatically transform a general-purpose LLM into a capable network diagnostic tool through SFT alone.
Figure~\ref{fig:results} provides a visual comparison across experimental conditions.

\subsection{Complexity and RAG Impact}

We analyze the results along two dimensions: problem complexity and the effect of retrieval-augmented generation.
The test set contains 252 single-parameter and 228 multi-parameter problems.
Table~\ref{tab:single_multi} presents the accuracy breakdown by problem complexity.
Both SFT models achieve higher accuracy on multi-parameter problems despite their inherent complexity, which can be attributed to the systematic elimination prompt being particularly effective for multi-parameter cases.

\begin{table}[t]
    \centering
    \caption{Accuracy (\%) by Problem Complexity}
    \label{tab:single_multi}
    \begin{tabular}{lcc}
        \toprule
        \textbf{Condition} & \textbf{Single-Param} & \textbf{Multi-Param} \\
        \midrule
        SFT (No-RAG) & 88.5 & 90.4 \\
        SFT (RAG) & 91.3 & 94.3 \\
        \bottomrule
    \end{tabular}
\end{table}

The RAG condition provides $\mathcal{S}(\text{OAI})$ context during mutation generation, enabling mutations that exercise specific code validation paths.
While RAG achieves higher overall accuracy (92.7\% vs.\ 89.4\%), the No-RAG condition achieves higher exact-match (85.4\% vs.\ 75.4\%).
This suggests that RAG-augmented traces teach the model to identify the correct parameter more reliably but sometimes express fixes in non-canonical formats learned from source code, resulting in score-2 matches rather than exact matches.
The comparable average scores (2.65 vs.\ 2.62) indicate that both approaches produce similarly capable models overall.
This is an important practical finding: the core synthetic data pipeline generates effective training data without requiring access to source code, lowering the barrier to adoption for operators who may not have source-level access to their network stack.

\section{Conclusions}
\label{sec:conclusion}

This paper presented a synthetic data generation pipeline and SFT approach for teaching LLMs to diagnose 5G gNB misconfigurations from error logs.
The pipeline systematically mutates OAI configurations, deploys them against a real testbed to capture error logs with known ground truth, and generates structured reasoning traces using three research-grounded prompt strategies (deductive chain, systematic elimination, and iterative self-refinement).
Experimental results on 480 held-out scenarios demonstrate that SFT on QwQ-32B improves exact-match accuracy from 13.8\% (zero-shot) to 85.4\%, with accuracy (score~$\geq 2$) reaching 92.7\% under RAG-augmented conditions.
Both RAG and non-RAG conditions produce comparably capable models, validating that the core synthetic data pipeline generates effective training data independent of source code access.

The pipeline is designed for generalizability: any operator with a testbed can replicate the approach by swapping OAI for their own network stack.
Future work will explore reinforcement learning techniques such as GRPO to further improve diagnostic accuracy.

\bibliographystyle{IEEEtran}
\bibliography{reference}
\newpage

\end{document}